\documentclass{article}
\usepackage{authblk}
\usepackage[top=2cm, bottom=2cm, left=2cm, right=2cm]{geometry}
\usepackage{fancyhdr}
\usepackage{setspace}
\usepackage{paralist}
\usepackage{hyperref}

\begin{document}

\title{An Observation About Passphrases: Syntax vs Entropy}
\author[1]{Eugene Panferov}
\date{} 

\maketitle

\begin{abstract}
\noindent On the premise that we are using passwords composed of multiple English words,
we argue that using syntactically correct passphrases has no significant impact on the security
in comparison to randomly arranged collections of words. We only analyze the contribution of the syntax itself.
A comparison to the other kinds of passwords is out of the scope.  
\end{abstract}
\vspace{2pc}

\section{Introduction}

It was suggested in \cite{cpsm} to use passphrases instead of ``traditional'' passwords, for multiple reasons,
including: sheer strength, memorability, and conforming to idiotic password creation policies without actually following detrimental recomendations of the policy authors.
This recomendation gives rise to a reasonable doubt: ``what if syntactically correct phrases are as weak as dictionary words in comparison to a random string of symbols?''
Indeed, syntax itself should weaken a passphrase, as it provides some ``predictability'' to the phrase. 
In present paper we address this problem, by comparing syntactically correct passphrases to random collections of words (which are considered sufficiently srong).

Before we begin, it is important to explain why and how we use \textit{entropy}.
Despite Shannon's entropy is shown (both experimentally \cite{weir} and theoretically \cite{cpsm}) to be NOT a measure of password strength,
and being massively misused by almost every ``computer scientist'' or ``security expert'',
it has some practical value: it plays the role of the most optimistic estimate for a password strength \cite{cpsm}.

Provided the attacker knows the defender's password choosing strategy, the password strength can not exceed the value of $2^{entropy}$.
It is easy to see. The password strength is defined in \cite{cpsm} as the expected value for the lenght of a guessing atack. 
The attacker's knowledge of the defender's strategy allows them to limit the search space to the defender's pool of passwords,
the cardinality of which is by definition $2^{entropy}$.

From now on we use the \textit{search space cardinality} as an expression of \textit{entropy},
as these terms are in direct correspondence to each other, and we do not need the logarithmic scale of \textit{entropy}.

Since we are going to investigate a negative impact on the password strength,
i.e. how much does the syntax of a passphrase \textbf{reduce} the password strength,
the search space cardinality is a good tool for the job, as it limits the strength from above.

\section{The Step One}

According to OED \cite{oxford}

The Second Edition of the 20-volume Oxford English Dictionary contains full entries for 171,476 words in current use, and 47,156 obsolete words.
To this may be added around 9,500 derivative words included as subentries. 
Over half of these words are nouns, 
about a quarter adjectives, 
and about a seventh verbs; 
the rest is made up of exclamations, conjunctions, prepositions, suffixes, etc. 
And these figures don't take account of entries with senses for different word classes (such as noun and adjective).

This suggests that there are, at the very least, 
a quarter of a million distinct English words, 
excluding inflections, and words from technical and regional vocabulary not covered by the OED, 
or words not yet added to the published dictionary, of which perhaps 20 per cent are no longer in current use. 
If distinct senses were counted, the total would probably approach three quarters of a million. 
Apparently we should ignore ``distinct cases'', thus the big picture is the following:

$$
total = 171476+47156 = T
$$
$$
nouns = T/2
$$
$$
adjectives = T/4
$$
$$
verbs = T/7
$$

the amount of Words (excluding exclamations, determiners, conjunctions and such)

$$
W = T/2 + T/4 + T/7 = 195207
$$
$$
nouns = (98/175)*W 
$$
$$
adjectives = (49/175)*W 
$$
$$
verbs = (28/175)*W 
$$

The search space for the random sequence of n words is: $S_{random}(n) = W^n$

We are to estimate the Search space $S(.)$ for a grammatically correct sentence by $S_{random}(n)$. To this end let us make a series of assumptions, that do not extend $S(.)$,
but simplify the estimation.

let's fix the sentence structure (as representing the worst case).

$$
subject - action - object - qualifier
$$

translated to the parts of speech, while omitting the grammar glue

$$
(adj,noun),(adverb,verb),(adj,noun),(adj,noun)
$$

Thus, for each position in the phrase the part of speech is known.
Let's assume that all adverbs are algorithmically derived from adjectives (so that we use the same pool for adverbs and adjectives), then 
the search space cardinaliy for the 8 keywords phrase is:

$$
T/4 * T/2 * T/4 * T/7 * T/4 * T/2 * T/4 * T/2 = T^8 / (512*7)
$$ 

or equivalently $W^8 / 7294$ which is significantly greater than $W^7$

\section{The Step Two}

We now generalize the estimate to any $n > 1$
and any fixed (singled out) structure of a phrase.

$S(n) = W^n * fraction_1 * fraction_2 * ..... * fraction_n$

where $fraction_i$ represents the fraction of the dictionary constituting the search space for the $i-th$ part of speech

In order to estimate these PoS fractions, we refer to \cite{foray} which lists the average frequencies of PoS in a set of 9 very different books:

$$
noun 19\%
$$
$$
verb 15\%
$$
$$
punctuation 14\%
$$
$$
preposition 13\%
$$
$$
determiner 10\%
$$
$$
pronoun 9\%
$$
$$
adverb 7\%
$$
$$
adjective 6\%
$$
$$
conjunction 4\%
$$
$$
other 3\%
$$
$$
symbol 1\%
$$

As we postulate ``practical'' bijection between adverbs and adjectives, we may unite them:
$adverb+adjective = 13\%$, then all nouns, adjectives, verbs total to: $19+15+13 = 47\%$ --
this sum represents the 100\% of our ``refined'' dictionary, therefore fractions of each of these 3 privileged PoS are:

$$
fraction_noun = 19/47
$$
$$
fraction_adj = 13/47
$$
$$
fraction_verb = 15/47
$$

In other words, provided we are reading a natural text
(any piece of meaningful English speech randomly selected from a corpus of all English texts),
every time we encounter a noun or adjective or verb,
there is a 19/47 chance this word is a noun,
13/47 chance it is an adjective, 15/47 chance it is a verb

Now, assuming that our password sentence is meaningful and grammatically correct,
we may expect PoS to appear in the sequence with the given frequencies.
So that we can \underline{weight} PoS fractions of the dictionary respectively,
obtaining an \underline{expected} value for the cardinality of the search space of a sentence of an arbitrary length n:

$$
S(n) = W^n * (19/47 * 98/175 + 13/47 * 49/175 + 15/47 * 28/175)^n
$$
$$
S(n) = W^n * ((19*98 + 13*49 + 15*28) /47/175)^n
$$
$$
S(n) = (0.35*W)^n
$$

Thus we may say that an average grammatically correct sentence with n keywords
is weaker than a random $n$ words sequence approx as if the dictionary is 3 times shorter.
In terms of the search space we can say that grammatical correctness reduces
the search space by the divider $3^n$.

\section{Conclusion}

Even if we fix the grammatical structure of a password sentence 
(modeling the worst choosing strategy)
and assume that all grammatical glue is known to the attacker
the search space for the sentence constitutes a large fraction 
of the search space $S_{random}(n)$ for the same $n$
namely $S_{random}(n)*(0.35^n)$
and guaranteed to be significantly higher than $S_{random}(n-1)$
up to the $n = 11$. In other words, it is enough to make your passphrase 1 word longer to compensate for the syntax weakness.

\end{document}